\documentclass[12pt,preprint]{aastex}

\shorttitle{55 Cancri: A Laboratory}
\shortauthors{Christodoulou \& Kazanas}

\begin{document}

\title{ 55 Cancri: A Laboratory for Testing \\ 
        Numerous Conjectures about Planet Formation}

\author{
Dimitris M. Christodoulou\altaffilmark{1,2} 
and 
Demosthenes Kazanas\altaffilmark{3}
}

\altaffiltext{1}{Math Methods, 54 Middlesex Tpke, Bedford, MA 01730.
			E-mail: dimitris@mathmethods.com}
\altaffiltext{2}{University of Massachusetts Lowell, Dept. of Mathematical
                 Sciences, Olney Hall, Room 428, \\ Lowell, MA 01854.
                        E-mail: Dimitris\_\_Christodoulou@uml.edu}
\altaffiltext{3}{NASA/GSFC, Code 663, Greenbelt, MD 20771.
			E-mail: Demos.Kazanas@nasa.gov}

\def\gsim{\mathrel{\raise.5ex\hbox{$>$}\mkern-14mu
             \lower0.6ex\hbox{$\sim$}}}

\def\lsim{\mathrel{\raise.3ex\hbox{$<$}\mkern-14mu
             \lower0.6ex\hbox{$\sim$}}}

\begin{abstract}

Five planets are presently believed to orbit the primary star of 55 Cnc,
but there exists a large 5 AU gap in their distribution between the two 
outermost planets. 
This gap has attracted considerable interest because it may contain
one or more lower--mass planets whose existence is not contradicted by 
long-term orbit stability analyses, in fact it is expected according to the 
"packed planetary systems" hypothesis and an empirical Titius--Bode relation
recently proposed for 55 Cnc.
Furthermore, the second largest planet is just the second farthest
and it orbits very close to the star. Its orbit, 
the most circular of all, appears to be 
nearly but not quite commensurable with the orbit of the third planet,
casting doubt that any migration or resonant capture of the inner planets 
has ever occurred and lending support to the idea of "in--situ" giant planet 
formation by the process of core accretion.
All of the above ideas will be tested in the coming years in this natural 
laboratory as more observations 
will become available. 
This opportunity presents itself in conjunction with a physical model that
relates the orbits of the observed planets to the structure of
the original protoplanetary disk that harbored their formation at the 
early stages of protostellar collapse. 
Using only the 5 observed planets of 55 Cnc, this model predicts that the 
surface density profile of its protoplanetary disk 
varied with distance $R$ precisely as $\Sigma (R)\propto R^{-3/2}$, 
as was also found for the minimum--mass solar nebula. Despite this similarity, 
the disk of 55 Cnc was smaller, heavier, and less rotationally supported 
than the solar nebula, so this system represents a different mode of 
multi--planet formation compared to our own solar system.

\end{abstract}




\section{Introduction}\label{intro}

\subsection{Observations}\label{obs}

Eighteen years of Doppler--shift measurements of the primary star in the
nearby binary system 55 Cancri (HD 75732; hereafter 55 Cnc) have gradually 
raised the number of orbiting planets from 1 (Butler et al. 1997), to 3 
(Marcy et al. 2002), to 4 (McArthur et al. 2004), and recently 
to 5 (Fischer et al. 2008). Although 
the 4 inner planets are closely packed together (their semimajor axes 
$a < 1$~AU), the orbits of all 5 planets are remarkably circular (see Table~1)
and presumably coplanar. The system exhibits astonishing regularity, 
reminiscent 
only of the planetary order in our solar system and unmatched by any other 
currently known multiple--planet extrasolar system (for reviews of orbital 
and dynamical characteristics of exoplanets, see Marcy et al. [2005] and 
Butler et al. [2006]). 
The most massive planet, 55 Cnc d with $a = 5.77$~AU and a minimum 
mass\footnote{Based on the inclination of 55 Cnc d, the inclination of the 
orbital plane appears to be $i=53^o$ (McArthur et al. 2004), which increases 
all masses by 25\%.}
of 3.835 Jupiter masses (Table~1), is the most distant gaseous 
giant known among exoplanets with well--defined orbits and, like Jupiter, 
it orbits beyond 5~AU from the star. Furthermore, there are no giant planets
in the area roughly between 1 and 5~AU, just like in our solar system. In fact,
no planets have been found to date in the 5~AU gap between the two
outermost planets, f and d, with semimajor axes $a=0.781$~AU and $a=5.77$~AU,
respectively.

\begin{table}[t]
\begin{center}
\begin{minipage}{7in}
\caption{}
\begin{tabular}{cccccc}
\multicolumn{6}{c}{\sc Characteristics of the Observed Planets} \\
\multicolumn{6}{c}{\sc Around 55 Cancri A (Fischer et al. 2008)} \\
\hline\hline
Index & Planet     & Orbital     & Semimajor  & Minimum                  & Orbital \\
 ~$i$ & Designation& Period      & Axis       & Mass                     & Eccentricity \\
      &            & $P_i$~(days)& $a_i$~(AU) & $M_{min, i}$~($M_J$)$^a$ & $e_i$   \\
\hline\hline
 1    &  e         &    2.817    & 0.038      & 0.034                    & 0.070     \\
 2    &  b         &   14.65     & 0.115      & 0.824                    & 0.014     \\
 3    &  c         &   44.34     & 0.240      & 0.169                    & 0.086     \\
 4    &  f         &  260.0      & 0.781      & 0.144                    & . . . $^b$\\
 5    &  d         & 5218        & 5.77       & 3.835                    & 0.025     \\

\hline\hline
\multicolumn{6}{l}{{\sc Notes}: }\\
\multicolumn{6}{l}{$(a)$~Minimum masses are given in Jupiter masses.} \\
\multicolumn{6}{l}{$(b)$~Undetermined value, but the data are consistent with $e_4=0$.} \\

\end{tabular}
\end{minipage}
\end{center}
\end{table}

Despite the similarities in orbital characteristics, the planetary system 
of 55 Cnc exhibits some conspicuous differences when compared to our solar 
system (Fischer et al. [2008] and Table~1): 
(a)~Three of the planets orbit too close to the star (within 0.24~AU),
that is closer than the orbit of Mercury around the Sun.
(b)~The second planet, 55 Cnc b ($a=0.115$~AU), is also the second largest
with a mass of $0.824-1.03$ Jupiter masses. This planet also shows the
lowest orbital eccentricity ($e=0.014$) among all 5 planets.
(c)~The 4 inner planets are all quite massive, with the innermost planet,
55 Cnc e ($a=0.038$~AU), having the smallest mass ($10.8-13.5$ Earth masses),
a value roughly comparable to the mass of Uranus. Most, if not all, of
these planets are expected to be gaseous giants in contrast to the 
terrestrial planets in our solar system.

The central star of 55 Cnc and the dynamics of its system have kept observers 
occupied for other reasons as well:
\begin{enumerate}
\item 
This star is the primary in a visual binary (separation $\sim$10$^3$~AU) 
and belongs to a group of $\sim$30 multiple stellar systems in which giant planets 
are seen orbiting very close to the primary stars (Eggenberger et al. 2004; 
Raghavan et al. 2006). In fact, the most massive inner planets tend to be found in 
multiple stellar systems and those with the shorter periods ($ P < 40$ days) tend to 
also have very low eccentricities. The three inner planets of 55 Cnc, with orbital
periods $P=2.82-44.3$ days and eccentricities $e=0.014-0.086$, all fit precisely 
this trend.
\item
The primary star of 55 Cnc has very high metallicity compared to the Sun 
($Z\simeq 2 Z_\odot$; 
Gonzalez \& Vanture 1998; Valenti \& Fischer 2005), a property that has 
been associated with a higher probability of forming massive protoplanetary cores 
and gaseous giant planets in the surrounding, also metal--rich, disk by the process 
of core accretion (Ida \& Lin 2004b; Bodenheimer \& Pollack 1986; Pollack et al. 1996). 
\item
It seemed for a time that an extended disk of dust was visible in the infrared,
but this result was negated
when additional infrared observations failed to detect a disk 
(Schneider et al. 2001; Bryden et al. 2006) and submillimeter observations indicated 
that the excess flux could be attributed to background sources (Jayawardhana et al. 2002). 
\item
The orbits of the two more massive of the inner planets, 55 Cnc b and c, 
appear to be close to but not quite in mean--motion resonance with a period 
ratio of nearly 3:1 (Marcy et al. 2002; Fischer et al. 2008).
This is the only known 3:1 apparent resonant pair,\footnote{Desort et al. (2008)
have just reported another possible 3:1 resonant pair in HD~60532 that may be 
confirmed within the next decade.} whereas in all the other known 
cases the planets appear to have been captured in 2:1 resonances (Butler et al. 2006).
\end{enumerate}

The picture that emerges from the observations of 55 Cnc is enigmatic.
The most important questions can be summarized as follows (Fischer et al. 2008): 
Why did this star end up with so many massive planets 
when 90\% of the observed stars do not have any giants at all? The similarities 
to our solar system seem to indicate that multiple--planet systems like 55 Cnc
(and $\mu$ Ara\footnote{The metal--rich star $\mu$ Ara (HD 160691, GJ 691) also
has a gaseous giant orbiting beyond 5~AU ($a=5.235$~AU) and 3 additional giants 
orbiting close to the star (Santos et al. 2004; Pepe et al. 
2007; and Table~5 below). Their distribution shows a large gap between 1.5 and 5.2~AU;
but the orbital eccentricities of the inner planets are comparatively larger 
($e=0.067-0.172$) than those in 55 Cnc (see also \S~\ref{muara}).}) 
should be common, but why have they not been observed in 
larger numbers?    
And if such multi--planet systems around metal--rich stars represent the new 
norm of planet evolution (Ida \& Lin 2004b), where are the systems with central 
stars chemically similar to our Sun?
One would certainly be reluctant to admit that our solar system is special
because of the inherent bias associated with such a proposition.
In addition, were the inner planets of 55 Cnc (and $\mu$ Ara) formed "in--situ"
on the observed orbits (Lissauer 1995; Bodenheimer et al. 2000), 
or are they the result of inward migration (Ward \& Hahn 2000; Lin et al. 2000;
Armitage et al. 2002;  Alibert et al. 2004, 2005a, 2006) 
that terminated conveniently (Marcy et al. 2000; Trilling et al. 2002;
Eggenberger et al. 2004)
to allow for the planets to survive and circularize their orbits?
And why are all the planetary orbits in 55 Cnc so nearly circular when large
eccentricities are quite common in exoplanets (Marcy et al. 2000, 2005) and 
it appears that resonant tidal interactions did not take hold
since an actual 3:1 commensurability was not established?
Finally, could there be smaller planets or an asteroid belt in the 5~AU gap
and, in any case, why was matter acquisition so disparate as to produce a series
of smaller planets surrounded by two very massive planets, 55 Cnc b and d?
Finding answers to even some of these questions can be a giant step toward
understanding planet formation and evolution. We will revisit the underlying
issues in \S~\ref{questions} below.

\subsection{Theoretical Models}\label{th}

The observations of 55 Cnc have motivated several empirical modeling efforts 
that attempted to explore the various compelling issues that arise from the 
impressive regularity of this planetary system, the existence of the 5~AU gap, 
and the unusual chemical properties of the parent star:

1.~{\it Metal--rich environment for planet formation}.---
Ida \& Lin (2004a, b, 2005) carried out Monte Carlo simulations of 
protoplanetary core growth, orbital migration, and gas acquisition.
Their model predicts that the probability of forming gas giants increases
rapidly with increasing metal abundances in the host stars and the
surrounding disks, which is consistent with the multitude of planets
observed in 55 Cnc and $\mu$ Ara. On the other hand, the physical and orbital 
properties of these planets are in disagreement with the main features
of this model, namely that there should not be planets with $10-100$ Earth masses
at $a=0.2-3$~AU and that short--period, Neptune--size planets should not be common
around such G-type stars as opposed to M dwarfs. This led Ida \& Lin (2005) to speculate
that the planets seen in 55 Cnc and $\mu$ Ara have not migrated to their present
locations but they were instead formed in--situ with the help of sweeping 
mean--motion or secular resonances. However, there is no evidence that gas-giant
migration or resonances ever played a role in the formation of these planets, 
and this adds to the mystery surrounding such well--ordered planetary systems.

2.~{\it Long--term orbital stability}.---
The stability of the planetary orbits in 55 Cnc has been investigated
independently by several groups: Raymond \& Barnes (2005) integrated the orbits
of the 3 most massive planets (b--d) for 100 Myr and found that the system could
afford to have more planets in stable orbits between planets c and d.
Fischer et al. (2008) integrated the orbits of all 5 planets for 1 Myr and found 
that planets b and c are not locked into a 3:1 mean--motion resonance.
Gayon et al. (2008) integrated the orbits of all 5 planets for 400 Myr and showed 
that the system represents a case of stable chaos in which, however, close encounters 
between planets are avoided. All of these models indicate clearly that 
the system is dynamically stable over long time scales ($\gsim$ 0.1 Gyr).

3.~{\it Apparent 3:1 mean--motion resonance}.---
The apparent 3:1 mean--motion resonance between the orbits of planets b and c 
in 55 Cnc was initially hailed as a new example of a low--order commensurability in
a "resonant system" (Barnes \& Quinn 2004), with 2:1 resonances providing
the only previous example in observed systems. Although Barnes \& Quinn (2004)
did not study 55 Cnc explicitly, their models showed that planets in resonant
systems have very narrow zones of stability and raised the possibility that
such systems are only marginally stable. Modeling of 55 Cnc was undertaken
by Zhou et al. (2008) who showed that planets b and c can be captured in a 3:1
resonance if they start with modest eccentricities ($e\sim 0.01-0.05$) 
and migrate slowly (for $\sim$0.1~Myr). The above models would indicate that the
arrangement of the inner planets in 55 Cnc is the result of special circumstances, 
an uncomfortable notion. Since we now know that planets b and c 
are not in resonance (Fischer et al. 2008; Gayon et al. 2008), a more palatable
conjecture is that no special migration has ever occurred in 55 Cnc and that its inner 
planets form a typical "interacting system" (Barnes \& Quinn 2004) in which
the planets are able to perturb one another and have broad zones of stability
available to them.

Modeling efforts and related conjectures concerning the apparent 3:1 resonance in 55 Cnc
demonstrate how difficult it is to simulate such complex systems because we do not
have a handle on the initial conditions; numerical models that start from arbitrary 
but convenient initial conditions are capable of producing "easily" just about 
all of the observed results. So it is not surprising that more recent modeling 
efforts (although they did not address 55 Cnc directly) now seem to provide evidence 
against 3:1 resonant captures of planets in general:
Pierens \& Nelson (2008) found that massive planets captured by a dominant,
Jupiter--size planet tend to end up exclusively in 2:1 or 3:2 mean--motion 
resonances, depending on their mass. 
Furthermore, Adams et al. (2008) and Lecoanet et al. (2008) found that turbulence 
in the host protoplanetary
disk can upset mean--motion resonances and thus resonant systems should be rarely
observed.

4.~{\it Packed planetary systems hypothesis}.---
The orbital stability calculations of Barnes \& Quinn (2004) and related simulations
by Barnes \& Raymond (2004) used massless test--particles in models of several known 
extrasolar multiplanetary systems in order to identify broad regions full of 
dynamically stable orbits and to predict the possible existence of additional, yet unseen
planets. These calculations were extended by Raymond \& Barnes (2005), Raymond et al.
(2006), and Barnes et al. (2008) to include various Saturn--mass and terrestrial--mass
particles in--between the orbits of the known planets in each simulated system.
Although it was easy to find stable orbits for additional planets with 
terrestrial or smaller masses in some of these systems, the main result of these 
numerical experiments came as a surprise: Many systems are on the edge of stability
and small changes to the values of $a$ or $e$ push them to become unstable. 
Following the ideas of Laskar (1996),
Barnes \& Raymond (2004) conjectured that the remaining apparently stable systems
are also packed with enough planets to be marginally stable. 
This is the packed planetary systems (PPS) hypothesis
and it predicts that stable systems may contain unseen planets that bring them
to the edge of stability.
The PPS models successfully predicted the existence of HD 74156 d (Barnes et al. 2008)
before its discovery (Bean et al. 2008) and proved to be consistent with the presence
of 55 Cnc f that was found (Fischer et al. 2008) orbiting at the inner edge of the 
predicted stable zone ($0.7$~AU $< a <$ 3.2~AU, $e < 0.2$; Barnes \& Raymond 2004).

The stable zone in 55 Cnc is too wide, as a result the 5--planet system is still 
not packed and there could be more undetected planets between the planets f and d
(Raymond et al. 2008). 
Specifically, the PPS models suggest the following interesting possibilities: 
(a)~A giant planet with up to Saturn mass may exist near 2~AU (the center of the stable 
zone) with $e\sim 0.08$ (Raymond \& Barnes 2005). 
(b)~Alternatively, several terrestrial planets with up to 0.63 Earth masses
may exist at 1.1--3.6~AU with $e\leq 0.36$ (Raymond et al. 2006). 
Although the possibility of having Earth--like planets in the "habitable zone" 
around 1~AU is intriguing, our own interest in this work is focused entirely on
the possibility of having a gas giant orbiting at $\sim$2~AU for the
reasons that we explain in \S~\ref{newmodel} below. 
The current observations of 55 Cnc do not rule out 
such a planet: At a distance of $\sim$2~AU, a giant planet could avoid 
detection if its minimum mass is smaller than 100 Earth masses 
(Fischer et al. 2008), i.e., approximately equal to or less than 
the mass of Saturn.

5.~{\it Empirical Titius--Bode relation}.---
Poveda \& Lara (2008) presented an exponential fit of the semimajor axes of
the 5 known planets in 55 Cnc. This empirical relation is akin to the 
Titius--Bode (TB) rule for the solar system, but it works only if there exists 
a hypothetical sixth planet at $a=2.08$~AU. Based on this result, 
Poveda \& Lara (2008) proposed that an unseen planet with $a\approx 2$~AU 
should be present in 55 Cnc. There is no physical basis for this prediction,
yet it is quite interesting that the empirical TB relation of 55 Cnc is in
complete agreement with the prediction of the PPS model for a gas giant at
the same orbital distance. 

Since the PPS model and the TB relation converge to the same prediction, 
they can both be tested by additional Doppler--shift measurements of the 
primary star of 55 Cnc. By extending their baseline, future observations 
should be able to search for an additional periodicity in the 5~AU gap 
of 55 Cnc and, irrespective of the outcome, the result can be 
pivotal for planet formation theories (see also \S~\ref{newmodel}.5).

6.~{\it The protoplanetary disk of 55 Cnc}.---
Fischer et al. (2008) made an attempt to estimate some key parameters
of the minimum--mass protoplanetary disk of 55 Cnc. Using a typical
surface density profile for the disk that varied with distance $R$ 
as $\Sigma (R)\propto R^{-3/2}$ and an estimate of the core masses 
of the 5 observed planets that took into account the excessive metallicity
of the system, they found a total mass of 0.031 $M_\odot$ out to 6~AU
and a surface density for the dust of 
$8.7 {\rm ~g} {\rm ~cm}^{-2}$ at 5~AU.
The dust-to-gas mass ratio was set to the nominal value of 1\% and
the mass estimate also took into account the intermediate inclination of the
orbital plane of the system ($i=53^o$, as derived by McArthur et al. [2004]
for 55 Cnc d).

The assumed $\Sigma (R)$ profile was first established for the minimum--mass 
solar nebula (MMSN) by Weidenschilling (1977) and Hayashi (1981) and it has been
commonly used ever since in calculations of this type.
Recently, this profile was investigated again in our
solar system (Davis 2005), in extrasolar systems (Kuchner 2004), and in
dusty circumstellar disks around T Tauri stars (Kitamura et al. 2002;
Andrews \& Williams 2007). Due to large systematic uncertainties, 
none of these investigations has been able to directly confirm 
or reject the original $\Sigma (R)$ relation, although all of them
provide useful hints about possible improvements of the "typical" 
surface density profile of the solar nebula and extrasolar
protoplanetary disks. For what follows, it is interesting to note that
the the model of Davis (2005) indicates that the original power--law 
$\Sigma\propto R^{k}$ with $k = -3/2$ tends to overestimate
the surface density within the central 1~AU of the solar nebula. 
Also, Kuchner (2004) finds a steeper profile with $k = -2$ for
the surface density of the typical minimum--mass "extrasolar nebula" (MMEN)
constructed by Weidenschilling's (1977) method from 26 exoplanets
found in multi--planet systems. The $k = -2$ power--law index results,
again, in an overestimate of the surface density within the central 1~AU
of the MMEN. The lack of a central core region, such as that suggested
for the MMSN (Lissauer 1987; Davis 2005) and for circumstellar disks
(Garaud \& Lin 2007), is the obvious reason for the 
excessive central densities in models that adopt a single power--law
density profile for the entire protoplanetary disk.

\subsection{A New Theoretical Model}\label{newmodel}

We are interested in 55 Cnc because this is the only extrasolar system
with so many planets in well--ordered, long--lived, circular orbits. 
We believe that the high degree of regularity seen in this system makes it 
an ideal laboratory for exploring the above--discussed issues and for 
testing the leading current hypotheses and models about planet formation 
in protostellar disks.

Our work starts with a new physical model that we formulated rigorously
based on exact solutions of the isothermal Lane--Emden equation with rotation
and that relates the observed planets in a well--ordered, multi--planet 
system to key physical properties of the protoplanetary disk that 
hosted their formation. 
First we applied this model successfully to the 11 largest planets of our solar 
system (Christodoulou \& Kazanas 2008, hereafter CK) and we proceed 
here to apply it to the 5 known planets of 55 Cnc. 

The results obtained from our modeling are tied to 
the theoretical models of \S~\ref{th} in the following ways:
\begin{enumerate}
\item
Our model uses all the planets in regular orbits irrespective of size, mass, 
and metal abundances. As such, it works equally well for solar--type systems 
like our solar system and for systems with different metallicities or with 
gas giants in short--period orbits like 55 Cnc. We do expect however that the
modeling results will have to be interpreted consistently with the physical and 
the chemical properties of the detected planets and their parent stars.
\item
It is important that the planetary orbits of 55 Cnc were found to be dynamically
stable over long time scales because our model associates such long--lived orbits
with local minima of the gravitational potential well in the midplane of the original
protoplanetary disk. This association does not hold for many odd systems 
in which one or two planets have migrated inward and are now seen very close 
to their parent stars. However, the regular spacing of the planets 
and the stability of 55 Cnc argue
against migration as a process that dominated its evolution.
\item
It is now clear that 55 Cnc b and c are not caught in a 3:1 mean--motion
resonance. This result also suggests that no significant migration has taken
place in the disk of 55 Cnc and it supports our picture of a well-ordered
system in which the planets formed in--situ by core accretion. In turn, 
such regular planetary orbits readily provide information about the locations 
of density enhancements (the local potential minima) in the disk during 
the planet formation stage. 
\item
The PPS models predict that there must exist at least one more planet in
the 5~AU gap of 55 Cnc and they will be tested by future observations 
for the first time in such a well--ordered, multi--planet, extrasolar system.
Our model is also not impervious to the possible existence
of an additional planet, but it does not necessarily need a 6$^{\rm th}$ planet
in order to work. As it predicts entirely different physical characteristics 
for the disk of 55 Cnc depending on whether 5 or 6 planets are used 
(see \S~\ref{properties}), our model too needs an observational resolution 
of this dilemma.
\item
Based on the work of CK, we now understand that the TB relation can be
accommodated within our approach, but the rule 
has no inherent physical significance for our solar system or for 55 Cnc. 
Since the TB relation of 55 Cnc necessarily predicts a 6$^{\rm th}$ 
planet at $a\approx 2$~AU, the absence of such a planet from the 5~AU gap 
will settle this issue without a doubt. Such an observational result will however
be detrimental to the PPS hypothesis as well. Alternatively, if an additional
planet is found in the 5~AU gap, then the doubts will persist, the PPS
hypothesis will persevere, and our model (and presumably models such as
those of Weidenschilling [1977] and Davis [2005]) will then show a clear
preference for a different radial density profile of the protoplanetary
disk of 55 Cnc (see \S~\ref{properties} for details).
\item
The attempt of Fischer et al. (2008) to estimate some of the properties 
of the minimum--mass protoplanetary nebula (MMPN)
of 55 Cnc is noteworthy by insufficient. The main drawback 
of these estimates is that the authors assumed a $\Sigma\propto R^{-3/2}$ 
profile for the disk without actually using Weidenschilling's (1977) 
method or any of the competing methods cited in \S~\ref{th}.6 above. 
Consequently, the particular density profile produced by our disk model
of 55 Cnc cannot be compared to this previous work. Some of the
parameter values obtained by Fischer et al. (2008) and by Kuchner (2004),
who did use Weidenschilling's method, can however be compared
to the results of our modeling, and we do so in \S~\ref{surface} below.
\end{enumerate}

\subsection{Outline}\label{outline}

In \S~\ref{model}, we describe in detail the new physical model of the 
midplane of the
protoplanetary disk of 55 Cnc and we derive the fundamental dynamical 
parameters of this disk (minimum mass, density profile, specific angular 
momentum, rotation frequency, and equation of state). 
We also examine the dynamical stability of the model and
we describe the effects that a possible 6$^{\rm th}$ planet will have 
if it is discovered orbiting at $a\approx 2$~AU, as expected by the PPS 
and TB models of the system (\S~\ref{th}).

In \S~\ref{discussion}, we interpret the results from our model and
we discuss the issues and the questions mentioned above that pertain 
to the impressive regular structure of this planetary system 
and its intriguing 5~AU gap. We also report briefly on our modeling
effort of the related multi--planet system $\mu$ Ara.

\section{Physical Model of the Protoplanetary Disk of 55 Cnc}\label{model}

\subsection{Basic Equations}\label{equations}

We are interested in the equilibrium structure of a rotating,
self--gravitating, isothermal disk of angular velocity $\Omega$,
mass density $\rho$, and sound speed $c_0$.
Following CK, we adopt cylindrical coordinates ($R, \phi, z$),
the assumption of axisymmetry ($\partial /\partial \phi = 0$),
and the assumption of cylindrical symmetry ($\partial /\partial z = 0$)
that is valid on the midplane of any disk. We cast all quantities in
dimensionless form as follows: We use the central density 
$\rho_0\equiv\rho (0)$ and the length scale of the disk
\begin{equation}
R_0 \equiv \frac{c_0}{\sqrt{4\pi G\rho_0}}\, ,
\label{length}
\end{equation}
where $G$ is the gravitational constant, and we define the normalized
radius as $x\equiv R/R_0$ and the normalized density as 
$\tau\equiv\rho /\rho_0$. In addition, we adopt a rotation profile of the form
\begin{equation}
\Omega (R) = \Omega_0 f(x)\, ,
\label{rot}
\end{equation}
where $\Omega_0\equiv\Omega (0)$ and the dimensionless function $f(x)$
is to be determined. Finally, we use $\Omega_0$ and the Jeans frequency
$\Omega_J\equiv \sqrt{2\pi G\rho_0}$ to define a dimensionless
rotation parameter of the form
\begin{equation}
\beta_0 \equiv \frac{\Omega_0}{\Omega_J}\, .
\label{beta}
\end{equation}
The equilibrium structure of the midplane of the disk is then described by
the isothermal Lane--Emden equation with rotation that takes the dimensionless form
\begin{equation}
\frac{1}{x} \frac{d}{dx} x \frac{d}{dx}\ln\tau \ + \ \tau \ = \ 
\frac{\beta_0^2}{2x}\frac{d}{dx}\left(x^2 f^2\right)\, .
\label{main1}
\end{equation}
In CK, we showed that any power--law density profile is an exact analytic
solution of this equation, provided that the rotation profile $f(x)$ is
also also determined self--consistently from eq.~(\ref{main1}).
Such "baseline" power--law profiles are particular solutions of the 
Lane--Emden equation, but they are unable to satisfy the central 
boundary conditions, namely that
\begin{equation}
\left\{ \begin{array}{c} 
         ~~\tau (0) \ = \ 1 \\
         \\
         \frac{\textstyle d\tau}{\textstyle dx} (0) \ = \ 0
         \end{array} \right\} \ .
\label{bc}
\end{equation}
They do however determine analytically the appropriate rotation profile 
$f(x)$ and the mean density profile irrespective of the applicable
(physical) boundary conditions.
The exact solutions can then be obtained by a simple numerical integration
that obeys the boundary conditions~(\ref{bc}), and they are bound to
oscillate permanently about the corresponding baseline solutions without
ever settling on to them (that would be a violation of the boundary conditions).
It is important to note that, although any monotonic power--law profile 
can be chosen as a baseline solution and it will produce a monotonic rotation
profile, the corresponding exact solution subject to the boundary 
conditions~(\ref{bc}) will still have to be oscillatory, creating thus a series
of local maxima and minima across the entire density profile.

\subsection{Baseline Model}\label{baseline}

For our baseline equilibrium model of the midplane of the MMPN of 55 Cnc, 
we adopt a composite analytic solution in which the mean density profile 
is a combination of a flat inner region (a "core") followed by a 
declining power--law:
\begin{equation}
\tau_{_{\textstyle base}} (x) \ = \ \beta_0^{\textstyle 2} \ \cdot 
         \left\{ \begin{array}{cc} 
         1 \ , & \ \ \ {\rm if} \ \ \ x \leq x_1 \\
         (x_1/x)^{\textstyle\delta} \ , & \ \ \ {\rm if} \ \ \ x > x_1
         \end{array} \right. \ ,
\label{den}
\end{equation}
where $x_1$ is the radius of the constant--density core region and $\delta > 2$.
Compared to the model of the solar nebula in CK, this model does not have
a flat outer region because no outer equidistant planets are observed in 55 Cnc.

The rotation pofile is determined next by substituting eq.~(\ref{den}) into
eq.~(\ref{main1}) and by solving the resulting differential equation:
\begin{equation}
f (x) \ = \ \left\{ \begin{array}{cc} 
           1 \ , \
         \ \ \ \ \ \ \ \ \ \ \ \ \ \ \ \ \ \ \ \ \ 
         \ \ \ \ \ \ \ \ \ \ \ \ \ \ \ \ \ \ \ \ \ 
         {\rm if} \ \ \ x \leq x_1 \\
             
         \sqrt{\frac{\textstyle 1}{\textstyle\delta - 2}\left[ 
               {\textstyle\delta}\left(x_1/x\right)^{\textstyle 2}
                        - {\textstyle 2}\left(x_1/x\right)^{\textstyle\delta}
                             \right]} \ ,
         \ \ \ {\rm if} \ \ \ x > x_1
         \end{array} \right. .
\label{den31}
\end{equation}
It is easy to show that this rotation law obeys the physical requirements that
$f(x)>0$ and $df/dx\leq 0$ everywhere for any choice of ~$\delta > 2$. 

The functions ~$\tau_{base} (x)/\beta_o^2$~ and ~$f(x)$~ are plotted in 
Figure~\ref{fig1} for ~$x_1=100$ ~and for various choices of the power--law 
index ~$\delta > 2$. We see that, as $\delta$ is increased, 
the rotation profile is not affected as much as the density profile.

\subsection{Solutions of the Boundary--Value Problem \\
            and Parameter Optimization}\label{exact}

The above composite equilibrium model is characterized by three free parameters: 
the core radius $x_1 > 0$, the rotation parameter $\beta_0\leq 1$, 
and the power--law index $\delta > 2$. 
The density profile (eq.~[\ref{den}]) of this baseline solution of the Lane--Emden 
equation~(\ref{main1}) is not capable of satisfying the physical boundary 
conditions~(\ref{bc}) at the center and it serves only as a mean 
approximation to the mean density of the corresponding physical model. 
The general form of the rotation law of the baseline (eq.~[\ref{den31}]) can, 
however, be adopted for the differential rotation $f(x)$ of the physical model 
as well. Then eq.~(\ref{den}) provides a prescription for the RHS of the Lane--Emden 
equation~(\ref{main1}) which can thus be written as
\begin{equation}
\frac{1}{x} \frac{d}{dx} x \frac{d}{dx}\ln\tau \ + \ \tau \ = \ 
\tau_{_{\textstyle base}} (x) \, .
\label{main2}
\end{equation}
This differential equation can be integrated numerically subject to the 
physical boundary conditions~(\ref{bc}). For all choices of the free parameters
\{$x_1>0$, $\beta_0\leq 1$, $\delta > 2$\}, the numerical solutions oscillate
about the baseline solutions~(\ref{den}). The local maxima of the density profile
correspond to minima of the gravitational potential well in the midplane of the
disk and they are ideal locations for the formation of planetary cores from the 
dust grains that also have to aggregate inside these wells. 

We construct a model of the MMPN of 55 Cnc by choosing the 3$^{\rm rd}$
planet ($a=0.240$~AU) from the star to set the physical scale of the model.
(Similar results are obtained when any of the inner 4 planetary orbits is used for
scaling.) Then the third density peak of the model (not counting the central peak at $x=0$)
is associated with a distance of 0.24~AU from the center,
and the exact solutions of the 
Lane--Emden equation~(\ref{main2}) are optimized by varying the
free parameters until the remaining density enhancements match as
closely as possible the observed semimajor axes of the remaining planetary orbits of
55 Cnc. The best--fit model is shown in Figure~\ref{fig2} and its density peaks
are listed in Table~2 along with the observed semimajor axes
given by Fischer et al. (2008).
The third density peak occurs at $x = 316.91$ in the best--fit model of 55 Cnc, 
implying that the length scale of the protoplanetary disk in the isothermal phase 
of its evolution was extremely small ($R_0 = 7.573\times 10^{-4}$~AU). 
This value is $\sim$30 times smaller than the length scale of our solar system 
($R_0 = 2.244\times 10^{-2}$~AU; CK) and indicates that the inner disk of 55 Cnc was 
denser than the solar nebula (see \S~\ref{properties} below).
Furthermore, there is no other density peak out to 100~AU, so this model predicts
no more planets beyond the orbit of 55 Cnc d.

\begin{table}[t]
\begin{center}
\begin{minipage}{7in}
\caption{}
\begin{tabular}{cccccc}
\multicolumn{6}{c}{\sc Density Peaks} \\
\multicolumn{6}{c}{\sc in the Best--Fit Model} \\
\multicolumn{6}{c}{\sc of the MMPN of 55 Cnc} \\
\hline\hline
Index & Planet     & Semimajor     & Peak       & Relative & Peak  \\
 ~$i$ & Designation& Axis          & Location   & Deviation& Density  \\
      &            & $a_i$ (AU)    & $d_i$ (AU) & (\%)     & $\tau (d_i)$ \\
\hline\hline
1     &  e         & 0.038         & 0.042      & ~~10.5   & $4.302\times 10^{-2}$ \\
2     &  b         & 0.115         & 0.091      & $-20.9$  & $8.076\times 10^{-3}$ \\
3     &  c         & 0.240         & 0.240      & ~~. . .  & $7.743\times 10^{-4}$ \\
4     &  f         & 0.781         & 0.862      & ~~10.4   & $3.561\times 10^{-5}$ \\
5     &  d         & 5.77          & 5.61       & $-2.8$   & $3.855\times 10^{-7}$ \\

\hline\hline

\end{tabular}
\end{minipage}
\end{center}
\end{table}

The relative deviations between the locations of the model peaks and the
planetary orbits are also listed in Table~2. The largest relative deviation 
($\sim$21\%) is observed for the second peak that corresponds to the orbit 
of planet b. This difference is still quite small ($\sim$0.02~AU) in absolute terms 
but it appears disproportionately large because our method of optimization
is skewed toward the inner planets for which it produces larger relative deviations
(see CK for more details).
The values listed in Table~2 cannot be improved individually because of the extreme
nonlinear nature of the problem. Only the value of $\beta_0$ can be optimized
independently of the other parameters by minimizing the deviation of the innermost
planet from the corresponding density peak. The strong coupling of the
remaining two parameters then demonstrates that fitting 5 planetary
orbits with 3 free parameters is not a simple matter, as nonlinearities do not 
generally allow for one-to-one relations between parameters and planetary positions.

\subsection{Physical Properties}\label{properties}

The parameters of the best--fit model determined by the optimizing algorithm
along with the scaling assumption that $d_3 = 0.24$~AU (see Table~2)
constitute a set of important dynamical parameters of the MMPN of 55 Cnc:
\begin{equation}
\left\{ \begin{array}{c} 
         \delta \ \ = \ 2.4708  \\
         \beta_0  \ = \ 0.1228  \\
         x_1 \ = \ 69.10 \\
         R_0 \ = \ 7.573\times 10^{-4}~~{\rm AU} \\
         R_1 \ \equiv \ x_1 R_0 \ = \ 5.233\times 10^{-2}~~{\rm AU} \\
         \end{array} \right\} \ .
\label{neb}
\end{equation}
The value of 
the power--law index $\delta\approx 2.5$ indicates that the density profile of the
differentially--rotating region of the disk declined with radius, on average, as
$R^{-2.5}$. This region extended outward beyond the core radius of $R_1\approx 0.05$~AU.
The low value of $\beta_0$ indicates that the rotation 
of the isothermal nebula was sufficiently slow to avoid nonaxisymmetric
instabilities (see \S~\ref{stability} below). 
The unusually small value of $R_0$, along with eq.~(\ref{length}) and a typical low
gas temperature, indicates that the nebula was very dense, about $\sim$10$^3$ times
denser than the solar nebula (see Table~3 below). 
So the picture that emerges from the best--fit model of 55 Cnc is
that of a relatively heavy and slowly rotating disk that is long--lived
and that can comfortably build massive planets over time by accumulating the solids
and then accreting the gas contained within each local potential well. 
 
The optimization procedure also finds some models of high quality in which
somewhat larger magnitudes of $\beta_0$ are compensated by smaller values of $x_1$.
The most different model gives us an idea about how shallow the area around the true 
minimum is in the three--parameter space of the model. This model is characterized by
the following values:
~$\delta = 2.4710$, ~$\beta_0 = 0.1583$, ~$R_0 = 9.839\times 10^{-4}$~AU, 
~$x_1 = 52.66$, ~$R_1 = 5.181\times 10^{-2}$~AU, ~and the relative deviations 
are smaller than 22\%.
However, the power--law index does not differ from 
~$\delta = 2.5$ ~by more than 1.2\% in any of the high--quality models.
Similarly, the variations of $R_0$ and $x_1$ always produce core radii
to within 1\% of $R_1=$0.052~AU in all of these models, so there is no doubt
that the core region of the MMPN of 55 Cnc was extremely small ($\sim$10 solar
radii).

The Lane--Emden equation that we solved can serve as a model of a
differentially--rotating disk supported by thermal pressure in the $z$--direction,
so we expect that the scale height from pressure support will be ~$H\propto R$ 
~down the radial density gradient.\footnote{The vertical scale height of a 
pressure--supported disk is
$$
H \ \sim \ \frac{c_0}{\Omega} \ \sim \ \frac{c_0}{v} R \ ,
$$
where $v$ is the rotation speed. In our models, $v$ becomes asymptotically flat 
when the density exhibits a power--law profile with $\delta > 2$ and $c_0$ is constant, 
leading to the approximate relation that ~$H\propto R$.} In this case, the 
value of ~$\delta\approx 2.5$ ~implies that the corresponding power--law index in the 
surface--density profile ~($\Sigma\propto R^{-\delta + 1}$) ~of the nebular disk 
was ~$k = 1 - \delta\approx -1.5$. This value is virtually identical to that obtained 
for the solar nebula by Weidenschilling (1977) who used an entirely different method 
and by CK who used the same optimization technique as in this work.

The conformance of these results does not hold if a 6$^{th}$ planet is assumed to exist
in the 5~AU gap of 55 Cnc. We have also optimized models that included a hypothetical
planet at $a = 2$~AU (as predicted by the PPS and TB models of 55 Cnc), and the best--fit
model of the six planetary orbits has the following parameter values:
~$\delta = 2.1174$, ~$\beta_0 = 0.1334$, ~$R_0 = 7.511\times 10^{-4}$~AU, 
~$x_1 = 56.02$, ~$R_1 = 4.208\times 10^{-2}$~AU, ~and the relative deviations 
are smaller than
19\%. These values are comparable to those in the best--fit model of the 5 observed
planets, except for $\delta$ whose smaller value produces a softer density gradient
to accommodate the hypothetical planet at 2~AU. This softer gradient can also 
accommodate one additional planet farther out, but the outermost density peak occurs 
at 20.56~AU which differs from $a\approx 15$~AU predicted for a hypothetical 
7$^{th}$ planet by the empirical TB relation of Poveda \& Lara (2008). This peak
is also not consistent with the recent PPS prediction (Raymond et al. 2008)
that an unseen outer planet could exist close to 9 or 10~AU, i.e. near the inner
edge of a wide zone of stability found beyond 55 Cnc d.
Thus, our model with hypothetical planets is still markedly different than 
the PPS and TB models of 55 Cnc.

Our modeling also delineates some of the fundamental physical characteristics
behind the mean 
density profile of the midplane of the disk of 55 Cnc. With the aid of our 
best--fit model, we can deduce substantial information concerning the structure 
and the dynamics of the nebular disk. In addition to the structural and rotational 
parameters discussed above, we can use our analytic baseline model in order to probe 
the dynamical state of the protoplanetary disk of 55 Cnc in its isothermal phase. 
Following CK, we calculate the relevant physical quantities for the MMPN of 55 Cnc
and we summarize our results in Table~3. 

\begin{table}[t]
\begin{center}
\begin{minipage}{7in}
\caption{}
\begin{tabular}{ccccc}
\multicolumn{5}{c}{\sc Physical Quantities of the Nebular Disks} \\
\multicolumn{5}{c}{\sc of 55 Cnc and the Solar System} \\
\hline\hline
Description & Quantity & & 55 Cnc Value $^a$ & Solar System Value $^a$ \\
\hline\hline
Equation of State 
& $c_0^2 / \rho_0 = 4\pi G R_0^2$ 
& $=$
& $1.08\times 10^{14}$
& $9.45\times 10^{16}$ 
\\
& $\overline{\mu}~\rho_0 / T$
& $=$ 
& $7.73\times 10^{-7}$
& $8.80\times 10^{-10}$
\\
\multicolumn{5}{l}{}\\
Central Density $^b$
& $\rho_0$
& $=$ 
& $3.30\times 10^{-6}$
& $3.76\times 10^{-9}$
\\
Jeans Frequency
& $\Omega_J = \sqrt{2\pi G\rho_0}$
& $=$ 
& $1.18\times 10^{-6}$
& $3.97\times 10^{-8}$
\\
Rotation Frequency
& $\Omega_0$
& $=$ 
& $1.44\times 10^{-7}$
& $1.65\times 10^{-8}$
\\
Rotation Period
& $2\pi /\Omega_0$
& $=$ 
& 1.4 \ yr
& 12 \ yr
\\
\multicolumn{5}{l}{}\\
\multicolumn{5}{l}{Core:~~$R\leq R_1, ~|z|\leq R_0$ $^c$}\\
\multicolumn{5}{l}{-----------------------------------}\\
Surface Density
& $\Sigma_0 = 2 R_0 \rho_0$
& $=$ 
& $74800$
& $2520$
\\
Mass
& $M_1$
& $=$ 
& $10^{-6} \ M_\odot$
& $10^{-4} \ M_\odot$
\\
Specific Angular Momentum
& $L_1/M_1 = \Omega_0 R_0^2 x_1^2/2 \beta_0^2$ 
& $=$ 
& $2.94\times 10^{18}$
& $7.19\times 10^{18}$
\\
\multicolumn{5}{l}{}\\
\multicolumn{5}{l}{Disk:~~$R\leq 6$~AU, ~$|z|\leq R_0$ $^c$}\\
\multicolumn{5}{l}{--------------------------------------}\\
Mass
& M 
& $=$ 
& $5\times 10^{-6} \ M_\odot$
& $4\times 10^{-4} \ M_\odot$
\\

\hline\hline
\multicolumn{5}{l}{{\sc Notes}: }\\
\multicolumn{5}{l}{$(a)$~cgs units apply unless stated otherwise. } \\
\multicolumn{5}{l}{$(b)$~$T = 10$~K (gas temperature) and 
$\overline{\mu} = 2.34 {\rm ~g} {\rm ~mol}^{-1}$ (mean molecular weight) assumed. }\\
\multicolumn{5}{l}{$(c)$~$R_0$ and $R_1$ are smaller in 55 Cnc by factors 
of $\sim$30 and $\sim$16, respectively. } \\

\end{tabular}
\end{minipage}
\end{center}
\end{table}

The corresponding results obtained by CK for the solar nebula 
are also listed in Table~3 for comparison purposes. 
Compared to the core of the solar nebula, the MMPN core of 55 Cnc was 
smaller ($R_1$ ratio $\sim$16),
denser ($\rho_0$ ratio $\sim$10$^3$),
heavier ($\Omega_J$ ratio $\sim$30), and
it was rotating $\sim$10 times faster. The disk
contained $\sim$100 times less mass\footnote{
The mass content of both models is considerably smaller than
that of the MMSN because we consider only the mass within
one length scale from the midplane of each disk (i.e.,
for $|z|\leq R_0$) and $R_0$ is extremely small in both cases.
The $z$--integration needs to be extended for about 1~AU away from the
midplane in order to produce masses $\sim$0.01 $M_\odot$.} 
within a 6-AU radius and only 2/5 of the 
angular momentum per unit mass. There is no doubt that
these two disks were different and their differences are
reflected in the distributions and the physical properties
of the various planets that they formed over time.

\subsection{Surface Density Profile}\label{surface}

Using the $\Sigma_0$ value of 55 Cnc from Table~3, the core radius 
$R_1=0.05233$~AU, and the power--law index $k\approx -1.5$, 
we can write the surface density profile of the MMPN of 55 Cnc 
for $R\geq R_1$ in the form
\begin{equation}
\Sigma (R) \ = \ 895 \left( \frac{R}{1 ~{\rm AU}}\right)^{-1.5} \ \ 
{\rm ~g} {\rm ~cm}^{-2} \, ,
\label{sigma}
\end{equation}
where $R$ is measured in~AU. This equation shows that
the surface density of the gas is
$\Sigma = 895 {\rm ~g} {\rm ~cm}^{-2}$ at $R = 1$~AU
and $\Sigma = 80 {\rm ~g} {\rm ~cm}^{-2}$ at $R = 5$~AU.
These values are 48\% of the corresponding values in the solar nebula.

The 1-AU value is in agreement with Kuchner's (2004) result
of $739 {\rm ~g} {\rm ~cm}^{-2}$, although he used just
the 3 planets b--d, the only ones known at the time. In particular,
planet f was not included in the model and this is the reason
that the 1-AU value was underestimated and the density profile 
turned out to be too steep with $k=-2.42$.
In general, undetected planets cause the main problem in this method,
and if such planets are included, they tend to soften the
density gradients and push Kuchner's average MMEN value of
$k=-2$ closer toward $-1.5$.

The 5-AU value corresponds to a typical surface density of solids of
$1 {\rm ~g} {\rm ~cm}^{-2}$ for a nominal value of 1\% for the
dust-to-gas mass ratio. This value is comparable to the low--end
values calculated by Fischer et al. (2008) using only the
planetary masses inferred from observations. Some additional estimates
by the same authors reach as high as $8.7 {\rm ~g} {\rm ~cm}^{-2}$.
Such estimates rely on uncertain assumptions about the masses of the
solid planetary cores and they are obviously too high, as they predict
much larger densities in the disk of 55 Cnc compared to those in the
solar nebula. But we believe that the opposite is true since our model
and Kuchner's (2004) model, following entirely different methodologies,
both predict significantly lower densities at 5~AU for the MMPN of 55 Cnc than 
the MMSN. We note that such low densities in the outer disk of 55 Cnc 
do not pose a problem for the formation of the massive planet d. 
As Fig.~\ref{fig2} shows, the half--width of the corresponding density 
peak is about 3~AU, and all the solids within this emormous potential well 
were bound to aggregate at the bottom and form the massive core of this planet.

\subsection{Dynamical Stability}\label{stability}

All composite power--law models with $\delta > 2$
described above are stable to axisymmetric perturbations because 
they satisfy the Rayleigh criterion:
The specific angular momentum $\Omega R^2$ is an increasing function 
of $R$ at all radii. This can be shown easily by using eq.~(\ref{den31})
to provide the function $f^2$ and then by proving that
$d(f^2x^4)/dx\geq 0$ for all $x$ and $\delta >2$.

The above models are also stable to nonaxisymmetric perturbations
because they do not rotate too fast. For sufficiently slow rotation,
the nonaxisymmetric modes are neutral and incapable of merging to
produce instability. This behavior is captured in the 
$\alpha$--parameter stability criterion (Christodoulou et al. 1995)
which for gaseous disk models can be written as 
\begin{equation}
\alpha \equiv \frac{1}{2}\left(\frac{\Omega_0}{\Omega_J}\right) \leq 0.35  \, .
\label{alpha}
\end{equation}
In our notation, this is equivalent to requiring that
$\beta_0\leq 0.7$ for stability, thus our best--fit model 
with $\beta_0\approx 0.12$ 
is far from the threshold of nonaxisymmetric instability.
A disk of dust would also be stable with this rotation law.
For an unisotropic disk of particles, the corresponding stability 
threshold switches from 0.35 to 0.25, and the requirement for stability 
then becomes $\beta_0\leq 0.5$.

\section{Summary and Discussion}\label{discussion}

\subsection{Summary}\label{sum}

In \S~\ref{intro}, we placed in proper context all the available
information about the multi--planetary system of 55 Cnc.
Our knowledge of this system comes from 18 years of Doppler--shift observations
of the central star
(Fischer et al. 2008; McArthur et al. 2004; Marcy et al. 2002; Butler et al. 1997)
and from theoretical modeling of the inferred planetary orbits
(Fischer et al. 2008;
Raymond et al. 2008; Poveda \& Lara 2008; Gayon et al. 2008;
Zhou et al. 2008; Raymond \& Barnes 2005; Ida \& Lin 2005; Kuchner 2004).
The metal--rich primary star of 55 Cnc has 5 massive planets in remarkably
circular orbits and, outside of our Sun, this is the only star
with so many planets in well--ordered, perfectly stable, nonresonant orbits
(Table~1).
The system does however exhibit a large 5~AU gap between its two outermost
planets and this vast empty region has sparked speculation that one or more 
additional planets may still remain undetected between 
planets f ($a=0.781$~AU) and d ($a=5.77$~AU).

Our contribution to the investigation of 55 Cnc is a new physical model
of the midplane of the protoplanetary disk in which the planets were formed.
This model was introduced in \S~\ref{newmodel} and it was described in
detail in \S~\ref{model}. It is based on exact solutions of the
isothermal Lane--Emden equation with rotation (eq.~[\ref{main1}]) subject to the
appropriate central boundary conditions (eq.~[\ref{bc}]). These 
solutions describe radial density profiles that are oscillatory by nature,
despite the fact that the corresponding rotation profiles are strictly monotonic
and well-behaved. The oscillations occur about an average (baseline) density
(eq.~[\ref{den}]) that is a combination of two power laws, a flat uniform
core and a decreasing outer section (\S~\ref{baseline}). 
Such baseline density profiles
are also exact intrinsic solutions of the Lane-Emden equation but they are
incapable of satisfying the central boundary conditions (see CK for more
details and for an application of the model to our solar system). The baselines 
do however obey the same rotation law (eq.~[\ref{den31}])
as the corresponding oscillatory solutions. 
Some typical baselines and their rotation profiles are shown in Fig.~\ref{fig1}, 
while a typical oscillatory solution and its baseline are shown in Fig.~\ref{fig2}.

The oscillatory density profile shown in Fig.~\ref{fig2}
represents the best--fit model for the midplane of the protoplanetary disk 
of 55 Cnc. The fit was obtained by nonlinear unconstrained optimization that
matches the orbital semimajor axes of the observed
planets to consecutive density maxima (or gravitational potential minima)
of the oscillatory profile 
(see \S~\ref{exact} and Table~2) and that uses the semimajor axis of the third
planet from the star to set the physical scale of the model. The dynamical parameters
of the best--fit model are listed in eq.~(\ref{neb}) and its physical
properties are summarized in \S\S~\ref{properties}--\ref{surface} and in Table~3.
The midplane of the disk of 55 Cnc was composed of a very small ($\sim$0.05~AU)
and very dense ($\sim$10$^{-6} {\rm ~g} {\rm ~cm}^{-3}$) uniform core
in slow rotation (period $\sim$1 yr), followed by a power--law gradient such that
the surface density varied with distance as $R^{-1.5}$ (eq.~[\ref{sigma}]) out 
to $\sim$6~AU.
This configuration is stable to both axisymmetric and nonaxisymmetric 
perturbations (\S~\ref{stability}).

\subsection{Discussion}\label{dis}

\subsubsection{The Two Modes of Multi--planet Formation}\label{modes}

The $R^{-1.5}$ dependence of the surface density profile of 55 Cnc is effectively
identical to that found by CK for the solar nebula. Despite this resemblance, the
detailed comparison between the two nebular disks shown in Table~3
indicates that the two models are generally dissimilar, and the differences seen 
in Table~3 are also reflected in the differing characteristics of the planetary systems 
that were created around these two stars. Specifically, the inner disk of 55 Cnc was denser
and heavier, and this explains why 4 massive planets were formed within the inner 0.8~AU.
These high densities are also responsible for the large growth of the second planet
from the star, 55 Cnc b, which is the most massive of the inner 4 planets. This planet was free
to grow in a high--density environment, while the area around the orbit of the first
planet could have been depleted by the protostar (just as Mercury's growth 
was severely limited by the protosun). On the other hand,
the same disk showed much smaller densities beyond 2~AU, and this explains why only
one massive planet managed to form at the bottom of the single, very wide potential well
outside of 5~AU.

\begin{table}[t]
\begin{center}
\begin{minipage}{7in}
\caption{}
\begin{tabular}{ccc}
\multicolumn{3}{c}{\sc The Two Modes of Multi--planet System Formation} \\
\hline\hline
 Physical      & Large Light Disks        & Small Heavy Disks       \\
 Property      & (Solar System) $^{a, b}$ & (55 Cancri A) $^{a, c}$ \\

\hline\hline
 & & \\
 Disk                                     &                      &              \\
 ---------------                          &                      &              \\
 Metallicity ($Z$)                        & $Z_\odot$            & $2 Z_\odot$  \\
 Radial Scale ($R_0$)                     & 0.02~AU              & 0.001~AU     \\
 Radial Size                              & 40--70~AU            & 6--10~AU     \\
 Core Size ($R_1$)                        & 1~AU                 & 0.05~AU      \\
 Core Density ($\rho_0$)                  & $4\times 10^{-9}$    & $3\times 10^{-6}$\\
 & & \\
 Rotation                                 &                      &              \\
 ---------------                          &                      &              \\
 \ \ \ \ \ $\beta_0$                      & 0.4                  & 0.1          \\
 Rotation Period ($2\pi /\Omega_0$)       & 12 \ yr              & 1 \ yr       \\
 Jeans Period ($2\pi /\Omega_J$)          & 5 \ yr               & 0.1 \ yr     \\
 & & \\
 Gas Surface Density                      &                      &              \\
 -----------------------------            &                      &              \\
 \ \ \ \ \                 $\Sigma_0$     & 2500                 & 75000        \\
 \ \ \ \ \       $\Sigma ({\rm 1 \ AU})$  & 2000                 & 1000         \\
 \ \ \ \ \       $\Sigma ({\rm 5 \ AU})$  & 200                  & 100          \\
 \ \ \ \ \ $\Sigma (R\geq R_1)\propto R^k$ , \ \ $k =$
                                          & $-1.5$               & $-1.5$       \\

\hline\hline
\multicolumn{3}{l}{{\sc Notes}: }\\
\multicolumn{3}{l}{$(a)$~cgs units apply unless stated otherwise. } \\
\multicolumn{3}{l}{$(b)$~Order-of-magnitude estimates based on the results of CK.} \\
\multicolumn{3}{l}{$(c)$~Order-of-magnitude estimates based on the results of this work.} \\

\end{tabular}
\end{minipage}
\end{center}
\end{table}

The planetary system of 55 Cnc and the solar system are both remarkably well--ordered
despite the differences in structure and dynamical characteristics and the differences
between their host stars. This comparison seems to indicate that there exist two modes
of planet formation that lead to stable, well--organized, multi--planet systems.
Based on our modeling and on the results of CK, we can describe these 
two extremes of multiple planet formation in the isothermal regime of protostellar 
collapse as follows (see also Table~4):
\begin{enumerate}
\item {\it Large Light Disks.}---Solar--type protoplanetary disks 
with $\sim$1~AU cores and moderate rotational support
(no more than $\sim$40\%--50\% of maximum rotation) 
that have core densities a few times above those required to form planets 
($\sim$10$^{-9} {\rm ~g} {\rm ~cm}^{-3}$; Lissauer 1993) are capable of forming rocky
planets in the first 2~AU and, due to the 
comparatively high densities in their outer regions, 
several gas giants beyond 5~AU as well. Such disks can extend for at least 40--70~AU
in the radial direction. Because their core densities are barely supercritical, these 
disks can form only terrestrial planets in the vicinity of the core radius.
\item {\it Small Heavy Disks.}---Metal--rich protoplanetary disks 
with $\sim$0.05~AU cores and little rotational support
(no more than $\sim$10\%--15\% of maximum rotation) 
that have core densities $\sim$1000 times above the critical value are capable of 
forming giant planets in the first 2~AU and, due to the substantial density drop 
in their outer regions, only one gas giant 
beyond 5~AU.\footnote{On the opposite end of chemical abundances, small and heavy,
metal--poor disks are not expected to form any gas giants beyond the inner few AU.
A possible example is HD~37124 (Vogt et al. 2005), a system with 3 giant
planets close to the star that was singled out by CK as the only other 
well--ordered multi--planet
system besides 55 Cnc. In HD~37124, there has been no
indication of a planet beyond the inner 3~AU.}
Such disks cannot extend for more than 6--10~AU (and if they do, their densities 
are too low to form giant planets).
Because their core densities are strongly supercritical, these disks can easily form
gas giants around 1~AU, so terrestrial planets are not
expected to be present in the habitable zone in this mode.
\end{enumerate} 

\subsubsection{Important Questions and Feasible  Answers}\label{questions}

Returning now to the questions posed in \S~\ref{obs}, we can outline 
some feasible answers suggested by the above results:
Most of the observed stars ($\sim$90\%) do not show any giant planets at all.
Some of them, especially those of solar type, may have planetary systems formed 
in large light disks (LLDs),
in which case their planets are not observable yet with the current resolution
because the inner planets have very low masses and the outer planets are too
distant from their stars. This is probably the reason why the solar system is
currently the only example of a system formed in an LLD.
On the other hand,
the inner giant planets formed in small heavy disks (SHDs) can be presently detected.
Besides the standard example of 55 Cnc, all other planets
found within 2--3~AU from their stars in nearly circular orbits are good candidates
for the SHD mode. This also includes $\mu$ Ara, HD~37124, and
the massive, short--period ($ P < 40$ days) planets 
with low eccentricities found in multiple stellar systems (see \S~\ref{obs}).

On the question of in--situ formation versus migration, the slow rotation found in
our models (Table~4) and the structure of the disks that show several local potential
minima strongly support the in--situ formation of planets by planitesimal aggregation
and gas accretion inside these potential wells. This does not mean that planetary
migration cannot occur at all, it just indicates that the multiple planets of 
well--ordered systems do not migrate. Migration may still be the dominant mechanism
in systems with planets in highly eccentric and/or irregularly spaced orbits. 
But even in such cases,
our modeling adds a new dimension to the migration problem that has not
been previously explored: The ability of a planet to migrate can no longer be
determined solely by the disk torques that redistribute the angular momentum of a
planetary orbit. That was done in the past (e.g., Lin et al. [2000], 
Ward \& Hahn [2000], and references therein)
while the density profile of the underlying disk was assumed to be
monotonically decreasing without any localized potential wells.
In the present setting in which the planets are trapped inside gravitational
potential minima, the energy content of each planet needs to be considered as well.
A planet orbiting inside a local gravitational potential well can migrate only 
if the interaction with the disk provides enough energy for it to leap out of its
well---otherwise the torqing of the planetary orbit will simply lead to stable 
oscillations and the development of some eccentricity insde this potential minimum.

On the question of the existence of additional planets in 55 Cnc, our model boldly
predicts that no other planet will be found in the future. In particular, we do not
believe that there is room in the 5~AU gap even for a small rocky terrestrial planet
or an asteroid belt because the potential minima of planets f and d reach into the 
gap (see Fig.~\ref{fig2}), and all debris must have been swept by these two massive planets.
This prediction is in contrast to the predictions of the PPS models (Raymond et al. 2008)
and the TB conjecture (Poveda \& Lara 2008) for the system, and the disagreement
can be resolved by future observations of 55 Cnc (see also \S~\ref{properties} above). 

The question of the impressive growth of inner planet 55 Cnc b ($a=0.115$~AU, 
mass $0.82-1.0$ Jupiter masses) was discussed briefly in \S~\ref{modes} above.
The density of the peak where this planet was formed (Table~2) is quite high
($\rho_2 = 8\times 10^{-3}\rho_0\approx 3\times 10^{-8} {\rm ~g} {\rm ~cm}^{-3}$).
Only planet e was formed in a higher--density environment in the SHD of 55 Cnc,
but this planet resides deep inside the core of the nebula and its gaseous
envelope could have easily been ablated by the protostar at a later time.
The situation is similar in the inner solar system,
although Venus is slightly less massive than the Earth.
The emerging paradigm, valid for both LLDs and SHDs, is that of a depleted
innermost planet followed by planets that are free to grow as much as the local
conditions in the disk would allow. This picture appears to be borne out 
also in another
SHD candidate, $\mu$~Ara, but with some caveats that we discuss below.

\subsubsection{The SHD of $\mu$ Ara}\label{muara}

Recent observations (Pepe et al. [2007] and Table~5) of the metal--rich star 
$\mu$~Ara ($Z\simeq 2 Z_\odot$; Santos et al. 2004) indicate 4 orbiting planets 
in a configuration very similar to that of 55 Cnc:
A single massive outer planet, $\mu$ Ara e, is separated from the 3 inner 
planets (c, d, and b) by a large empty region between 1.5 and 5.2~AU.
The minimum mass of the innermost planet is comparable to that of the innermost planet 
in 55 Cnc, so this planet is also a low--mass (depleted) giant while its two neighbors
have minimum masses comparable to the mass of Jupiter. 

\begin{table}[t]
\begin{center}
\begin{minipage}{7in}
\caption{}
\begin{tabular}{cccccc}
\multicolumn{6}{c}{\sc Characteristics of the Observed Planets} \\
\multicolumn{6}{c}{\sc Around $\mu$~Ara (Pepe et al. 2007)} \\
\hline\hline
Index & Planet     & Orbital     & Semimajor  & Minimum                  & Orbital \\
 ~$i$ & Designation& Period      & Axis       & Mass                     & Eccentricity \\
      &            & $P_i$~(days)& $a_i$~(AU) & $M_{min, i}$~($M_J$)$^a$ & $e_i$   \\
\hline\hline
 1    &  c         &    9.6386   & 0.091      & 0.03321                  & 0.172   \\
 2    &  d         &  310.55     & 0.921      & 0.5219                   & 0.0666  \\
 3    &  b         &  643.25     & 1.497      & 1.676                    & 0.128   \\
 4    &  e         & 4205.8      & 5.235      & 1.814                    & 0.0985  \\

\hline\hline
\multicolumn{6}{l}{{\sc Note}: }\\
\multicolumn{6}{l}{$(a)$~Minimum masses are given in Jupiter masses.} \\

\end{tabular}
\end{minipage}
\end{center}
\end{table}

Although the eccentricities of its planetary orbits are larger than those of 55 Cnc
(see Tables~1 and 5),
$\mu$~Ara appears to be a good example of an SHD system. The size of the
system is also comparable to that of 55 Cnc ($\sim$6~AU), so we expected our 
best--fit model to produce parameter values similar to those of 55 Cnc.
We also thought that, by analogy, a PPS model of the system would find a stable zone
in the 3.7~AU gap between planets b and e, and a TB model would predict a planet 
in the same gap. Such models are not yet available, but the TB relation of
Poveda \& Lara (2008) can be easily applied to the observed semimajor axes.
It turns out that the TB models are not particularly good fits to the data with 
or without a hypothetical 5$^{th}$ planet in the 3.7~AU gap.

The planetary orbits of $\mu$~Ara shown in Table~5 cannot be matched to the 
density maxima
in any of our models. The reason for this failure is that
planet d is too close to planet b, and this leaves an inconspicuous 0.8~AU gap 
in the distribution between planets c and d. A hypothetical planet is needed
in this gap to restore the regular spacing of the inner planets. 
This, of course, does not mean that an undetected planet necessarily exists in
the 0.8~AU gap, it just implies that a density maximum is needed in that area
of the model for the distribution to be typical of a well--organized SHD system.
By the same token, some models allow for yet another density maximum
interior to the orbit of planet c, at $\sim$0.05~AU. Such experimental models
are characterized by different parameter sets, so a "best--fit model" cannot 
be determined, unless additional planets are discovered in $\mu$~Ara by future
observations.

\subsubsection{Remarks on Surface Density Profiles}\label{sdp}

A surface density profile of the form ~$\Sigma (R)\propto R^{-1.5}$~
is applicable outside the central core region
of the protoplanetary disk of 55 Cnc and the solar nebula
(Table~4). The presence of a roughly uniform core is strongly suggested 
by both theory and observations (see \S~\ref{th}.6), and our models
suggest that LLDs and SHDs can be distinguished by their core sizes
($\sim$1~AU and $\sim$0.05~AU, respectively). On the other hand, we do not
know why both of the nebular disks in Table~3 show the $R^{-1.5}$
trend, although this fact---that we independently obtained the same result 
(as opposed to two different power laws) from two different planetary 
systems---argues in favor of our modeling assumptions.

At this point, we do not have any more evidence of the ubiquity of the
$R^{-1.5}$ profile among LLDs and SHDs, but we can certainly argue
that this density power law is more relevant to planet formation
than the profiles from steady--state viscous and irradiated accretion disks
(Kitamura et al. [2002], Andrews \& Williams [2007], Garaud \& Lin [2007],
and references therein). The isothermal phase of protostellar collapse
that our models address
is an extended intermediate phase of the whole process during which
the magnetic fields have diffused away, the central core mass is a 
minute fraction of the solar mass, matter continues being
deposited onto the disk from above and below, most of the gas in the disk 
is molecular, and all the energy produced is efficiently radiated away 
(see Tohline [2002] for a recent review). At this stage, the evolution
of the gas is not driven by viscosity or by ionization because the gas
is transparent to radiation; the gas responds fully to the effects of
self--gravity and (non--Keplerian) rotation, and this is why it is
very encouraging to know that our modeled disks are dynamically stable
(\S~\ref{stability}). Theoretical calculations of viscous, magnetized,
or irradiated accretion disks are not applicable in this regime
of protostellar collapse, while observations of thermal radiation from
T Tauri stars probe the disks long after the isothermal phase is over,
the gas has become opaque to its own radiation, and the grown protostars have
imposed nearly Keplerian rotation profiles onto the disks.
Therefore, there is no reason to expect that the surface 
density profiles should have power--law indices of $k=-1$ or softer
(seen in T Tauri disks; Kitamura et al. 2002; Andrews \& Williams 2007) 
during the isothermal phase of collapse; and if the planetary results
of Kuchner (2004) give any indication about extrasolar systems, we should
expect instead $k$--indices between $-1.5$ and $-2$.



\newpage

\section*{FIGURE CAPTIONS}

\figcaption{Analytic density and rotation profiles of the composite 
equilibrium models described by eqs.~(\ref{den}) and~(\ref{den31})
for $x_1=100$ and $\delta =$ 2.1, 2.5, and 2.9. Both profiles are 
uniform in the core region with $x\leq x_1$. 
\label{fig1}
}

\figcaption{Equilibrium density profile in the midplane of the protoplanetary 
disk of 55 Cnc. 
The composite model described in \S~\ref{baseline} (eq.~[\ref{den}], dashed line) 
has been adopted for the RHS of the Lane--Emden equation~(\ref{main2}) and this 
differential equation has 
been integrated numerically subject to the physical boundary conditions~(\ref{bc}). The 
resulting exact solution (solid line) has been fitted to 55 Cnc so that its 
density maxima (dots) correspond to the observed semimajor axes of the planetary orbits 
(open circles).
Frame (a) shows the radial distance ~$d$ ~on a linear scale out to ~$d=7$~AU. ~Frame (b) 
shows the same radial distance on a logarithmic scale out to ~$\ln d = 3$ ~($d=20$~AU).
~The nonrotating analytic solution $\ln\tau (d) = -2\ln (1 + d^2/8R_0^2)$ from CK
(dash-dotted line) is also shown for reference. The numerical solution tracks 
closely the nonrotating solution near the center, but eventually it has to turn 
around to approach the (dashed) baseline solution; and this sets off the oscillations 
in the physical density profile.
\label{fig2}
}

\newpage

\begin{figure}[t]
\vskip 7.5truein
\includegraphics{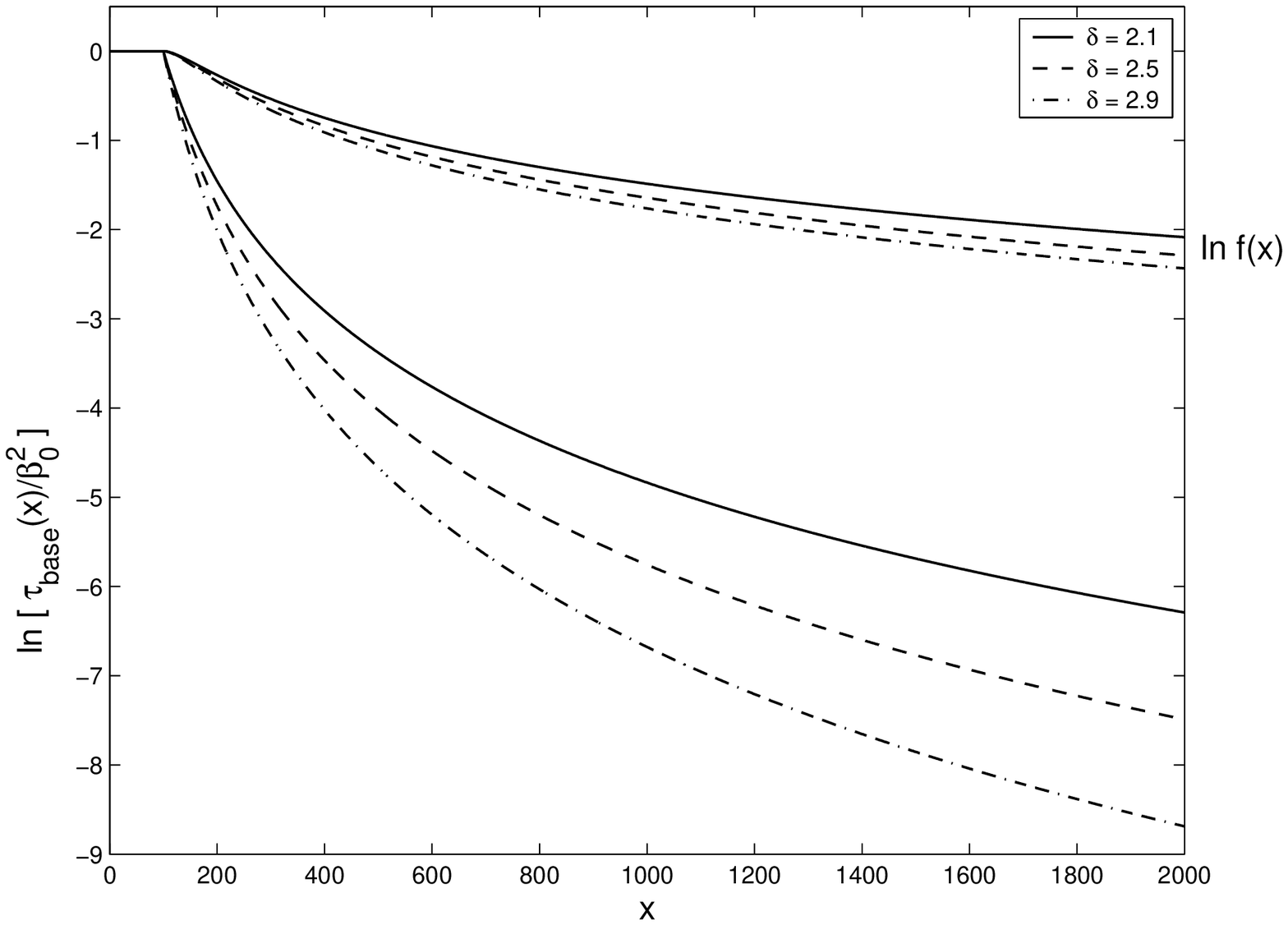}
{\bf FIGURE 1}
\end{figure}

\newpage

\begin{figure}[t]
\vskip 2.5truein
\includegraphics{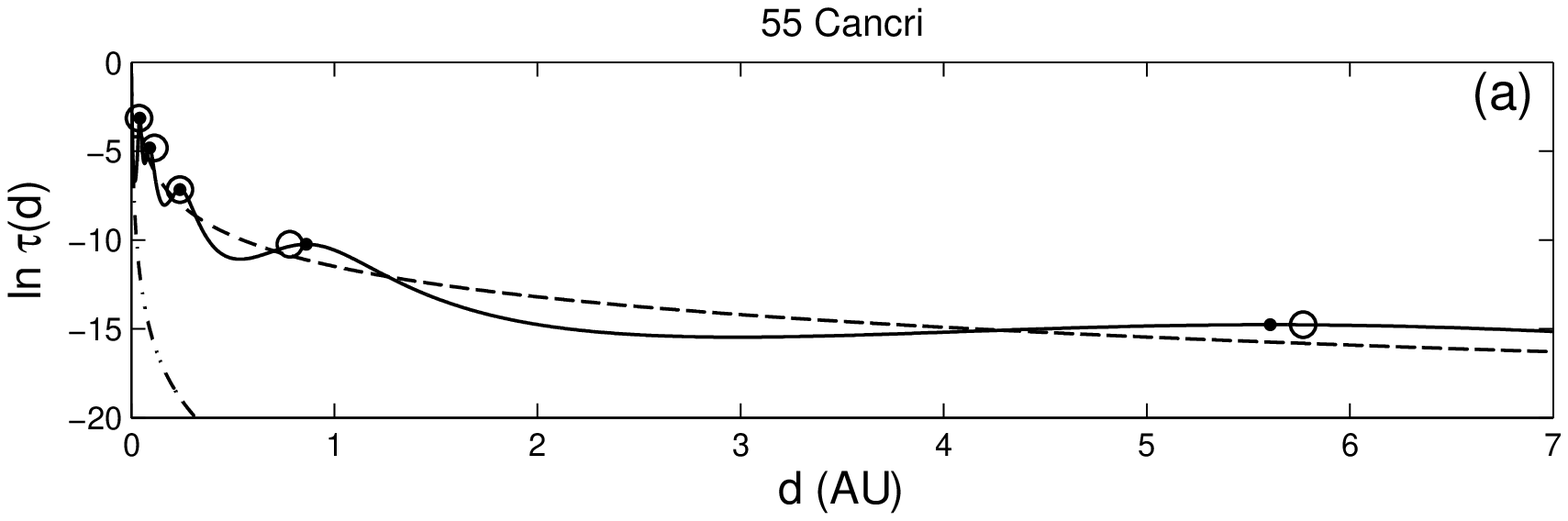}
\vskip 2.5truein
\includegraphics{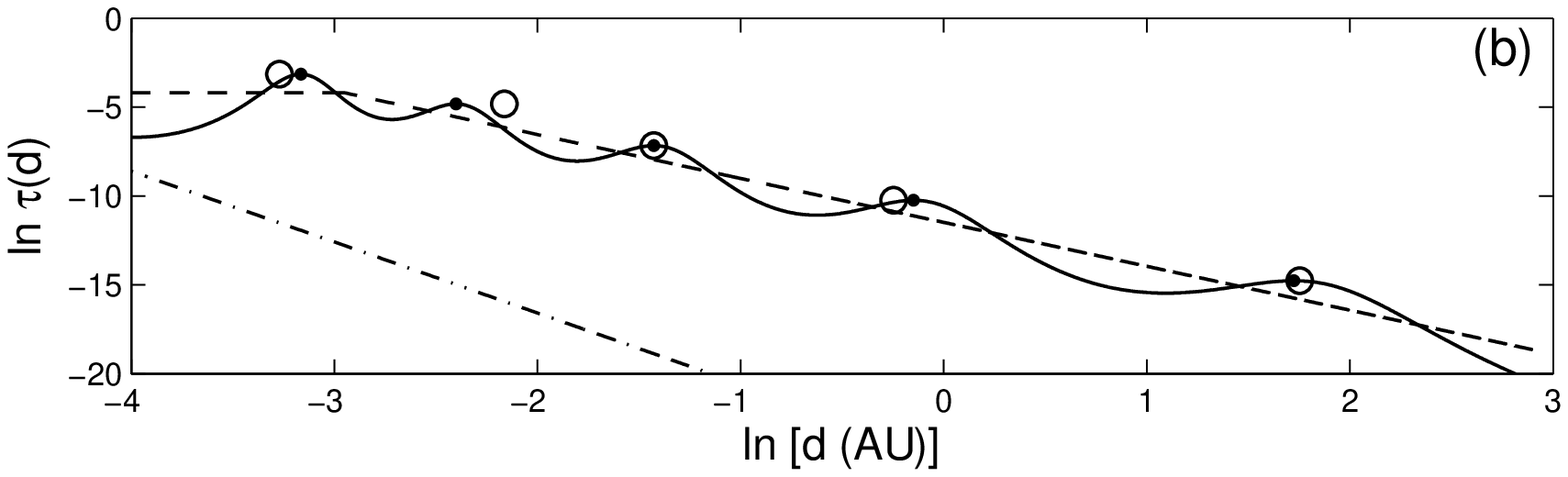}
{\bf FIGURE 2}
\end{figure}

\end{document}